\documentclass[preprint,12pt]{elsarticle}

\usepackage{amsmath}
\usepackage{amssymb}
\usepackage{amsfonts}
\usepackage{graphics}
\usepackage{graphicx}
\usepackage{amscd}
\usepackage{amsfonts}
\usepackage{color}
\usepackage{braket}
\usepackage{comment}
\usepackage{hyperref}

\usepackage{ulem}

 \usepackage[utf8]{inputenc} 

\biboptions{numbers,sort&compress}

\usepackage{etoolbox}
\makeatletter
\patchcmd{\ps@pprintTitle}
  {\let\@oddhead\@empty}
  {\def\@oddhead{\mbox{}\hfill \footnotesize{KUNS-2797, RIKEN-iTHEMS-Report-20
}}}
  {}{}
\makeatother

\newcommand{\n}{\nonumber}
\newcommand{\la}{\langle}
\newcommand{\ra}{\rangle}
\newcommand{\eref}[1]{(\ref{#1})}

\newcommand{\f}[2]{\frac{#1}{#2}}

\allowdisplaybreaks[1]

\begin{document}

\begin{frontmatter}

\title{Obtaining the sphaleron field configurations with gradient flow}

\author{Yu Hamada}
\address{Department of Physics, Kyoto University, Kyoto 606-8502, Japan}
\ead{yu.hamada@gauge.scphys.kyoto-u.ac.jp}

\author{Kengo Kikuchi}
\address{Department of Physics, Kyoto University, Kyoto 606-8502, Japan\\
RIKEN iTHEMS, Wako, Saitama 351-0198, Japan}
\ead{kengo@yukawa.kyoto-u.ac.jp}

\begin{abstract}
 We propose a formalism to obtain the electroweak sphaleron,
 which is one of the static classical solutions,
 using the gradient flow method.
 By adding a modification term to the gradient flow equation,
 we can obtain the sphaleron configuration as a stable fixed point of the flow in the large flow time.
 Applying the method to the $SU(2)$-Higgs model (the Weinberg angle $\theta_W$ is $0$) in four dimensions,
 we obtain the sphaleron configuration whose energy coincides with previous studies numerically.
 We can also show that the Chern-Simons number of the solution has a half-integer.
\end{abstract}


\end{frontmatter}

\section{Introduction} 
\label{sec:introduction}

One of the most important problems that remains unanswered by the Standard Model (SM) of particle physics
is the baryon asymmetry in the universe (BAU). Various scenarios have been proposed in particle physics and cosmology to explain the BAU,
electroweak baryogenesis \cite{Kuzmin:1985mm},
leptogenesis \cite{Fukugita:1986hr},
and baryogenesis via neutrino oscillations \cite{Asaka:2005pn}.
(For reviews of electroweak baryogenesis, see, e.g., \cite{Morrissey:2012db,Trodden:1998ym,Rubakov:1996vz,Cohen:1993nk}.)

The sphaleron\cite{Klinkhamer:1984di} plays a quite important role in most of the scenarios
to generate a sufficient amount of baryon asymmetry.
It is an unstable solution of equation of motions
and a saddle point of the action with only one tachyonic mode around it.
In the electroweak theory, particularly, it is on top of an energetic barrier
between two vacuum configurations that have different topology,
which are labeled by integers called the Chern-Simons number.
It is important to understand the properties of the sphaleron deeply
for giving precise predictions on the BAU.

One of the conventional methods to obtain the sphaleron is
the so-called minimization and maximization procedure\cite{Manton:1983nd}.
This method works quite well for simple cases,
such as the case of $\theta_W =0$ (i.e., $SU(2)$-Higgs theory)
or sufficiently small $\theta_W \simeq 0$\cite{Klinkhamer:1984di}.
However, it is difficult to carry out such a procedure for finite $\theta_W \neq 0$
because the sphaleron solution is no longer spherically symmetric.

In this paper, we give a much simpler and easier formalism to obtain the sphaleron configuration in the electroweak theory using the gradient flow method. 
In general, the gradient flow method has played an important role in various research areas of physics.
The flow equation is constructed by the gradient of the action, which gives a classical solution at large flow time. This method is used as a traditional way to solve the system, for example, one of them is called a relaxation method to solve the nonlinear system. On the other hand, in gauge theories, the gradient flow in Yang-Mills theory has been the focus of much attention\cite{Luscher:2010iy, Luscher:2011bx}. The flow has specific properties for the finiteness of correlation functions, and there are a lot of studies including the lattice field theory.

Here we consider the gradient flow to find the sphaleron solution.
The idea is based on the work in Ref.~\cite{Chigusa:2019wxb}.
In the reference, the authors proposed the method to determine the bounce configuration
for false vacuum decay using the gradient flow.
The modification term was added to the flow equation to make the bounce a stable fixed point.
(See also Refs.~\cite{Sato:2019axv,Ho:2019ads} for related works.)
Based on the work, we propose a new method to obtain the sphaleron solutions.

The most important point is how to choose the modification term appropriately.
In this paper, we give the term as a variation of the Chern-Simons number with respect to the fields.
The modified flow equation in this manner gives the sphaleron configuration
as a stable fixed point of the equation in the large flow time.
Applying the method to the $SU(2)$-Higgs model in four dimensions,
we give the modified flow equation of the model.
Solving the flow equation numerically,
we indeed obtain the sphaleron configuration that agrees with previous studies in a good precision.
We also show that the Chern-Simons number of the solution converges with a half-integer along the flow.
These results suggest the validity of our method to obtain the electroweak sphaleron
without any fine-tuning of the initial configurations.

This paper is organized as follows.
In Sec.~\ref{sec:formulation}, we give a brief review on the modified gradient flow method given in Ref.~\cite{Chigusa:2019wxb} from the standpoint of finding sphalerons and propose a new method to obtain the sphaleron.
In Sec.~\ref{sec:model}, we apply this method to the $SU(2)$-Higgs theory concretely.
Solving the flow equation numerically, we show that the method to obtain the sphaleron works well in Sec.~\ref{sec:numerical}.
And a summary and discussion is given in Sec.~\ref{sec:summary}.


\section{Formulation} 
\label{sec:formulation}
We present a simple formulation to obtain the sphaleron configuration in gauge theories,
using the gradient flow method.
Our formulation is based on the work in Ref.~\cite{Chigusa:2019wxb},
where the bounce configurations are mainly considered.
(See also Refs.~\cite{Sato:2019axv,Ho:2019ads} for related works.)
We apply their method for obtaining sphalerons.
To be specific, we concentrate on static configurations in $(3+1)$ dimensions,
but our argument can be extended to any dimensions.

\subsection{Brief review on modified gradient flow method}
The authors of Ref.~\cite{Chigusa:2019wxb} proposed the modified gradient flow equation 
to obtain the bounce solutions in scalar theories.
Their method can be easily extended to other saddle point solutions, such as the sphalerons.
In this subsection, we give a brief review on the work of Ref.~\cite{Chigusa:2019wxb} 
from the standpoint of finding sphalerons.

Let $\Phi_A(x,s)$ be a multiplet describing all fields in the model
and $A$ run over the number of degrees of freedom of the fields.
We consider the sphaleron configuration as an example of saddle point solutions,
which is denoted by $\Phi_A^{sph.}(x)$.
The modified gradient flow equation proposed in Ref.~\cite{Chigusa:2019wxb} is given by
\begin{equation}
  \partial_s \Phi_A(x,s) = \mathcal{F}_A (x,s) - \tilde{\beta}~ \mathcal{G}_A (x,s),\label{025758_25Nov19}
\end{equation}
where $x$ is the Euclidean three-dimensional coordinate and $s$ is a fictitious time called the flow time. 
The first term of the r.h.s.~$\mathcal{F}_A(x,s) $ is defined by
the gradient of the energy functional,
\begin{equation}
 \mathcal{F}_A(x,s) \equiv - \frac{\delta E[\Phi]}{\delta \Phi_A(x,s)}.
\end{equation}
The second term of the r.h.s., which is called a modification term, 
is introduced to make the sphaleron a stable fixed point.
Here $\tilde{\beta}$ is defined by
\begin{equation}
\tilde{\beta} \equiv \beta \f{\braket{\mathcal{G}|\mathcal{F}}} {\braket{\mathcal{G}|\mathcal{G}}},
\end{equation}
with $\beta$ being a constant larger than unity, and the inner product of two sets of functions is defined as
\begin{equation}
 \braket{f|f'} = \int d^3 x~ \sum_A f^\dagger_A f'_A.
\end{equation}

We define the fluctuation operator (or the quadratic curvature) as
\begin{equation}
 \mathcal{M}_{AB}[\Phi^{sph.}] \equiv \left. \f{\delta^2 E[\Phi]}{\delta \Phi_A \delta \Phi_B} \right|_{\Phi \to \Phi^{sph.}},\label{103606_4Feb20}
\end{equation}
which has one negative eigenvalue and other positive ones.
In order for the flow equation, Eq.~(\ref{025758_25Nov19}), to converge to the sphaleron,
the function $\mathcal{G}_A(x,s)$ should be chosen to be proportional to the eigenfunction with the negative eigenvalue \cite{Chigusa:2019wxb}.

\subsection{Choice of $\mathcal{G}$}
In order for the flow equation, Eq.~(\ref{025758_25Nov19}), to work, 
it is most important to choose $\mathcal{G}_A$ appropriately.
Obviously, this is not an easy task because we do not know the precise expression of the negative eigenfunction in general.
In this subsection, we propose a new method on 
how to choose the function $\mathcal{G}_A$ appropriately to obtain the sphalerons.

We define $\mathcal{G}_A$ as the gradient of the Chern-Simons number,
\begin{equation}
 \mathcal{G}_A(x,s) \equiv \frac{\delta N_{CS}[\Phi]}{\delta \Phi_A(x,s)}.\label{181134_27Nov19}
\end{equation}
Since we consider $(3+1)$-dimensional gauge theories, 
\begin{equation}
 N_{CS} = \frac{-1}{16 \pi^2} \int_{\mathbb{R}^3}  \mathrm{tr}\left( AF - \frac{2}{3} A^3 \right),\label{NCS}
\end{equation}
and $\mathcal{G}_A$ can be expressed as
\begin{equation}
 \mathcal{G}_A =
  \begin{cases}
   (8\pi^2)^{-1} ~\epsilon^{ijk}F_{jk}^a & (\Phi_A = A_i^a )\\
   0 &  (\Phi_A= \mathrm{other~matter~ fields}),
  \end{cases}
\end{equation}
where $i,j,k$ are the three-dimensional indices and $a$ is the adjoint index.

\begin{figure}[htbp]
 \centering
\includegraphics[width=0.72\textwidth]{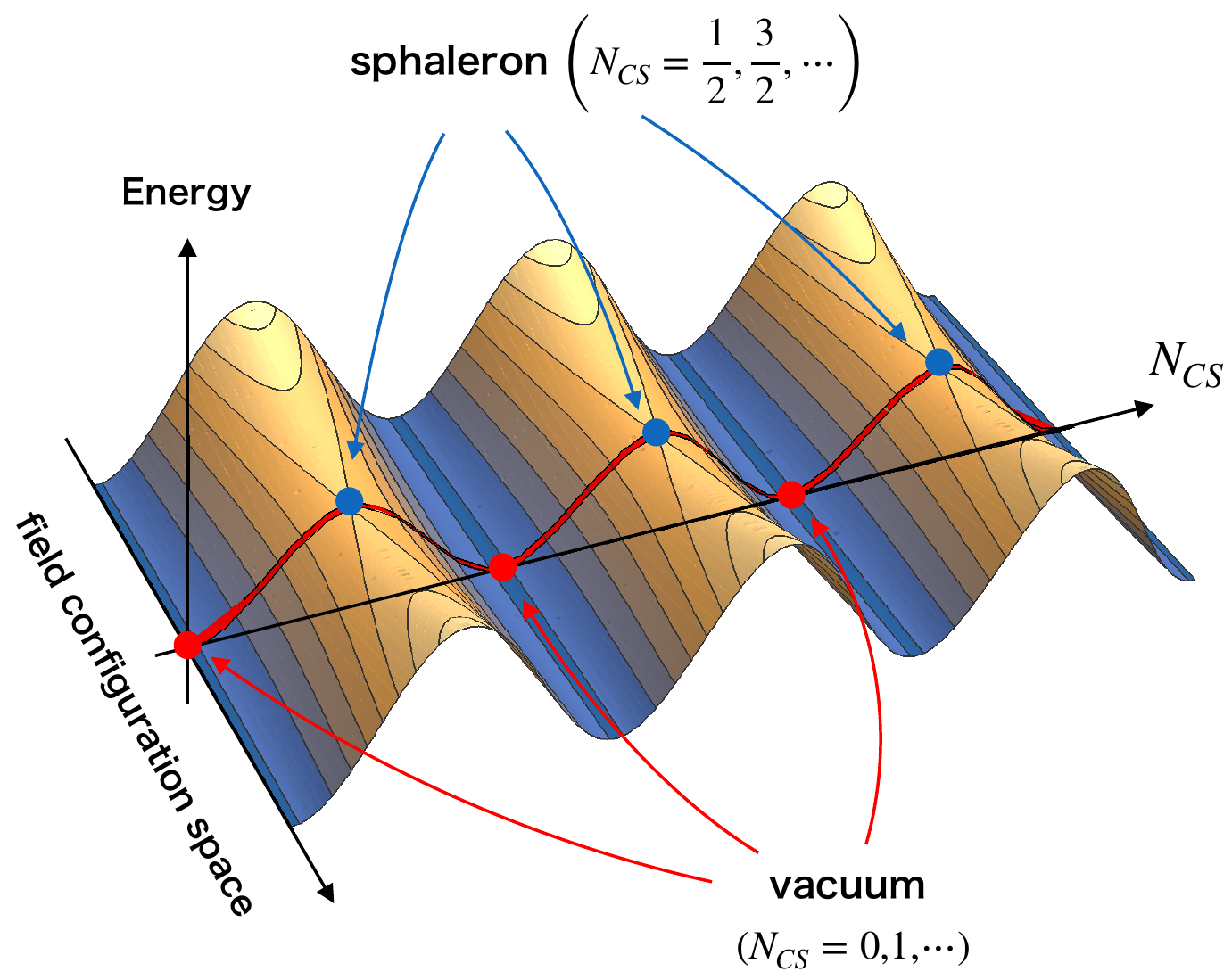}
 \caption{The naive picture of the sphaleron in the configuration space.
 Red points are the infinitely degenerated vacua.
 Blue points, which are saddle points of the energy, are the sphaleron solutions. }
\label{131820_6Jan20}
\end{figure}

We explain why the choice Eq.~(\ref{181134_27Nov19}) is appropriate,
i.e., it is proportional to the negative eigenfunction of the operator Eq.~(\ref{103606_4Feb20}).
As is well known, the sphaleron is a saddle point
on top of the energetic barrier 
separating two topologically different vacua
with integral Chern-Simons numbers $N_{CS}=n, n+1$ $(n\in\mathbb{Z})$.
Because of the reflective symmetry between the two vacua,
the sphaleron has the half-integer Chern-Simons number 
\footnote{
This statement would not be correct if one considers the deformed sphalerons, 
which are another kinds of sphalerons and have non-half-integer values of $N_{CS}$.
In this paper, we do not consider such a case.
}
: $N_{CS}=n+ \f{1}{2}$.
The naive picture of the sphaleron in the configuration space is shown in Fig.\ref{131820_6Jan20}.
One of the axes indicates the value of $N_{CS}$, and energy barriers separating degenerated vacua are orthogonal to the axis of $N_{CS}$. The sphalerons are located on the saddle points on the barriers.
According to the picture, the positive eigenmodes of $\mathcal{M}$
are along the barrier and do not change the Chern-Simons number,
while the negative eigenmode (the direction to roll down the barrier) changes the Chern-Simons number.
Thus, we naively guess that
the negative eigenfunction is the steepest direction changing the Chern-Simons number,
i.e., the gradient of $N_{CS}$ : $\delta N_{CS}/\delta \Phi$.
Therefore, the choice Eq.~(\ref{181134_27Nov19}) would be appropriate.

Using the flow equation Eq.~(\ref{025758_25Nov19}) with Eq.~(\ref{181134_27Nov19}), we can obtain the sphalerons quite easily.
All that needs to be done is to evolve an initial configuration by the flow equation Eq.~(\ref{025758_25Nov19})
keeping some appropriate boundary conditions.
In the following sections, we explicitly show numerical results that strongly suggest that
the method with the choice Eq.~(\ref{181134_27Nov19}) works well for the sphalerons.

\section{The Model} 
\label{sec:model}

In this section, we actually apply the formulation to the $SU(2)$-Higgs model, which is well known for the existence of the sphaleron\cite{Klinkhamer:1984di,Manton:1983nd,Dashen:1974ck,Yaffe:1989ms}, and show that our method works well to obtain the sphaleron.

\subsection{SU(2)-Higgs Model }

We give a brief review of the $SU(2)$-Higgs model in four dimensions based on Ref.~\cite{Yaffe:1989ms}. 
The action of the model in four dimensions is given by 
\begin{eqnarray}
S=\frac{1}{g^2}\int d^4x\left[-\frac{1}{2}\mathrm{tr}\left(F_{\mu\nu}F^{\mu\nu}\right)+(D_\mu\Phi)^\dagger(D^\mu\Phi) -\frac{\lambda}{g^2}\left(\Phi^\dagger\Phi-\frac{1}{2}g^2v^2  \right)^2\right].\label{4dimaction}~~~~~~
\end{eqnarray} 
where the covariant derivative is $D_\mu \equiv \partial_\mu+ A_\mu$, $\Phi$ and $A_\mu=A_\mu^a (\sigma^a /2i)$ are a scalar field and a $SU(2)$ gauge field, respectively. $\sigma^a$ means Pauli matrices. The field strength is defined by $F_{\mu\nu}=[D_\mu, D_\nu]$. Here we use the Minkowski metric diag$(+,-,-,-)$. 

Imposing the spherically symmetric condition, the model reduces to two-dimensional $U(1)$ gauge theory, 
which is coupled to two complex scalar fields.  
The ansatz for spherically symmetric configurations for $A_\mu$ and $\Phi$ are described by  \cite{Ratra:1987dp} 
\begin{eqnarray}
A_0(x)&=&\frac{1}{2i}\left\{ a_0(r,t)\hat{x}_j \sigma^j \right\},\\
A_i(x)&=&\frac{1}{2i}\left[ \left\{ f(r,t)-1\right\}\frac{e_i^1}{r}+h(r,t) \frac{e_i^2}{r}+a_1(r,t)e_i^3  \right],\\
\Phi(x)&=&\left\{\mu(r,t)+i\nu(r,t)\hat{x}_j\sigma^j \right\}\xi,
\end{eqnarray}
where  $\hat{x}$ and $\xi$ are a radial unit vector and a two component complex unit vector, respectively. 
The coordinates $\{e_i^k\}$ are defined by  
\begin{eqnarray}
e_i^1&=&\epsilon_{ijk}\hat{x}^k\sigma^j,\\
e_i^2&=&(\delta_{ij}-\hat{x}_i \hat{x}_j)\sigma^j,\\
e_i^3&=&\hat{x}_i \hat{x}_j\sigma^j.
\end{eqnarray}
For the sake of convenience, 
we introduce two complex scalar fields $\chi\equiv f+ih, \phi\equiv\mu+i\nu$ 
and a field strength for the two-dimensional gauge field $a_\mu$ as $f_{\mu\nu}=\partial_\mu a_\nu-\partial_\nu a_\mu$.
Hereinafter, the subscripts $\mu$ and $\nu$ run 0 and 1. 
Then the action in four dimensions \eref{4dimaction} reduces to one in two dimensions as
\begin{eqnarray}
&&S=\frac{4\pi}{g^2}\int dt dr \bigg\{\frac{1}{4}r^2 f_{\mu\nu}f^{\mu\nu}+|D_\mu\chi|^2
+\frac{1}{2r^2}\left(|\chi|^2-1 \right)^2+r^2|D_\mu\phi|^2\n\\
&&\hspace{10mm}-\mathrm{Re}\left(\chi^*\phi^2\right)+\frac{1}{2}\left(|\chi|^2+1\right)|\phi|^2
-\frac{\lambda}{g^2}r^2\left(|\phi|^2-\frac{1}{2}g^2v^2 \right)^2 \bigg\},
\end{eqnarray}
where $D_\mu \chi=(\partial_\mu-i a_\mu)\chi$ and $D_\mu\phi=(\partial_\mu-i a_\mu/2)\phi$.

The finite energy configurations should satisfy the following conditions: 
\begin{eqnarray}
\chi\rightarrow e^{i\omega},~~~\phi  \rightarrow e^{i\frac{\omega}{2}}\frac{gv}{\sqrt{2}},~~~a_\mu\rightarrow \partial_\mu\omega \label{161907_14Feb20}
\end{eqnarray}
as $r\rightarrow\infty$, and
\begin{eqnarray}
|\chi|\rightarrow1,~~~ D_\mu\chi\rightarrow0, ~~~\chi|\phi|^2\rightarrow\phi^2\label{161830_14Feb20}
\end{eqnarray}
as $r\rightarrow0$.
The first two conditions and the third one in Eq.~\eqref{161830_14Feb20} are required to remove the singularities of the field strength
and the gauge current at short distance, respectively.

We can also confirm that the Chern-Simons number in four dimensions given in Eq.~\eref{NCS} reduces to one in two dimensions using ansatz as follows:
\begin{eqnarray}
N_{\mathrm{CS}}=-\frac{1}{2\pi}\int dr \epsilon^{01}\left[a_1-\mathrm{Im}\partial_1\chi+\frac{1}{2i}\left\{\chi^*(D_1\chi)-(D_1\chi)^*\chi  \right\}\right].
\end{eqnarray}

\subsection{Modified gradient flow equation}

We give the modified gradient flow equation of the $SU(2)$-Higgs model 
using the formulation in Sec.~\ref{sec:formulation}. 
Since we are interested in a static solution, 
we impose a static condition that 
the derivatives with respect to the time and the time component of the gauge field vanish. 
Under this condition, we still have a residual gauge symmetry generated by a time-independent gauge function $\omega(r)$,
which is completely fixed by the radial gauge condition $a_1=0$.
Thus, the modified gradient flow equations for the sphaleron are described by 
\begin{eqnarray}
\frac{\partial \chi}{\partial s}&=&-\frac{\delta E}{\delta \chi^*}-\beta\frac{\la \mathcal{G} | \mathcal{F}\ra}{\la \mathcal{G} |\mathcal{G}\ra }\frac{\delta N_{\mathrm{CS}} }{\delta \chi^*},\label{maineq1} \\
\frac{\partial \phi}{\partial s}&=&-\frac{\delta E}{\delta \phi^*}-\beta\frac{\la \mathcal{G} | \mathcal{F}\ra}{\la \mathcal{G} |\mathcal{G}\ra }\frac{\delta N_{\mathrm{CS}} }{\delta \phi^*} \label{maineq2},
\end{eqnarray}
where
\begin{eqnarray}
\frac{\delta E}{\delta \chi^*}&=&\left[-\partial_1^2+\frac{1}{r^2}\left(|\chi|^2-1  \right)+\frac{1}{2}|\phi|^2 \right]\chi-\frac{1}{2}\phi^2,\\
\frac{\delta E}{\delta \phi^*}&=&\left[\partial_1 r^2 \partial_1+\frac{1}{2}\left(|\chi |^2+1\right)  -\frac{\lambda}{g^2}r^2\left(2|\phi|^2-g^2 v^2 \right)  \right]\phi-\chi\phi^*,\\
\frac{\delta N_{\mathrm{CS}}}{\delta \chi^*}&=&2i\frac{\partial \chi}{\partial r},\\
\frac{\delta N_{\mathrm{CS}}}{\delta \phi*}&=&0,
\end{eqnarray}
and
\begin{eqnarray}
\la \mathcal{G} | \mathcal{F}\ra&=&\int dr \left\{\left(\frac{\delta N_{\mathrm{CS}}}{\delta \chi} \right)^*\frac{\delta E}{\delta \chi}+\left(\frac{\delta N_{\mathrm{CS}}}{\delta \chi^*} \right)^* \frac{\delta E}{\delta \chi^*}
\right\},\\
\la \mathcal{G} | \mathcal{G}\ra&=&\int dr \left(\left| \frac{\delta N_{\mathrm{CS}}}{\delta \chi}\right|^2+\left| \frac{\delta N_{\mathrm{CS}}}{\delta \chi^*}\right|^2
 \right).
\end{eqnarray}
Equations \eref{maineq1} and \eref{maineq2} give a stable fixed point
we are looking for as the sphaleron.  
In the next section, we show that the configuration is indeed the sphaleron 
in the $SU(2)$-Higgs model numerically.

\section{Numerical analysis} 
\label{sec:numerical}
In this section, we show numerical results of the $SU(2)$-Higgs sphaleron
obtained by the flow equations \eqref{maineq1} and \eqref{maineq2}.

\subsection{Setup}
To perform numerical calculations, 
we take a length unit such that $v=1$ and a spatial lattice 
as $r= i\times \Delta r$ with $i=1,2,\cdots, N$ and $\Delta r=5.0 \times 10^{-2}$.
The size of the system $L$ is given by $L\equiv N\times \Delta r$.
To avoid numerical instability, a lattice spacing for the flow time $\Delta s$ should satisfy
$\Delta s \leq (\Delta r)^2=2.5 \times 10^{-3}$,
so that we take $\Delta s = 1.5\times 10^{-3}$.

For any flow time $s$, the functions $\chi(r,s)$ and $\phi(r,s)$ should satisfy 
the conditions Eqs.~(\ref{161907_14Feb20}) and (\ref{161830_14Feb20}) with $a_\mu(x,s)=0$.
In numerical calculations in this paper, therefore, we adopt the following boundary conditions:
\begin{equation}
\begin{cases}
 \partial_r\chi (0) = 0,~ \phi (0) = 0 & (\text{at the origin})\label{163412_3Feb20} \\
   \chi (L) = 1,~ \phi (L) = gv/ \sqrt{2} & (\text{at large distances})
 \end{cases}
\end{equation}
or equivalently, 
\begin{equation}
 \begin{cases}
 f'(0)=h'(0)=0,~ \mu(0)=\nu(0)=0  & (\text{at the origin}) \label{184233_9Dec19} \\
 f(L)=1,~  \mu(L)= gv/\sqrt{2},~ h(L)=\nu(L)=0 & (\text{at large distances})
 \end{cases}
\end{equation}
where $L$ should be sufficiently large comparing to the size of the sphaleron.
All boundary conditions except for $\phi(0)=0$ in Eq.~(\ref{163412_3Feb20})
follow from the conditions Eqs.~(\ref{161907_14Feb20}) and (\ref{161830_14Feb20}) immediately.
The reason of $\phi(0)=0$ is that $\phi(r)$ should be an odd function with respect to $r$.
\footnote{
If $\phi(r)$ were an even function, it must satisfy $\partial_r \phi(0)=0$ at the origin.
Under these boundary conditions, we obtain the following unique solution to the equation of motions:
$\chi(r)=1, \phi(r)= gv / \sqrt{2}$,
which is the trivial vacuum solution.
}

\subsection{Result}
\begin{figure}[htbp]
\centering
 \includegraphics[width=1\textwidth]{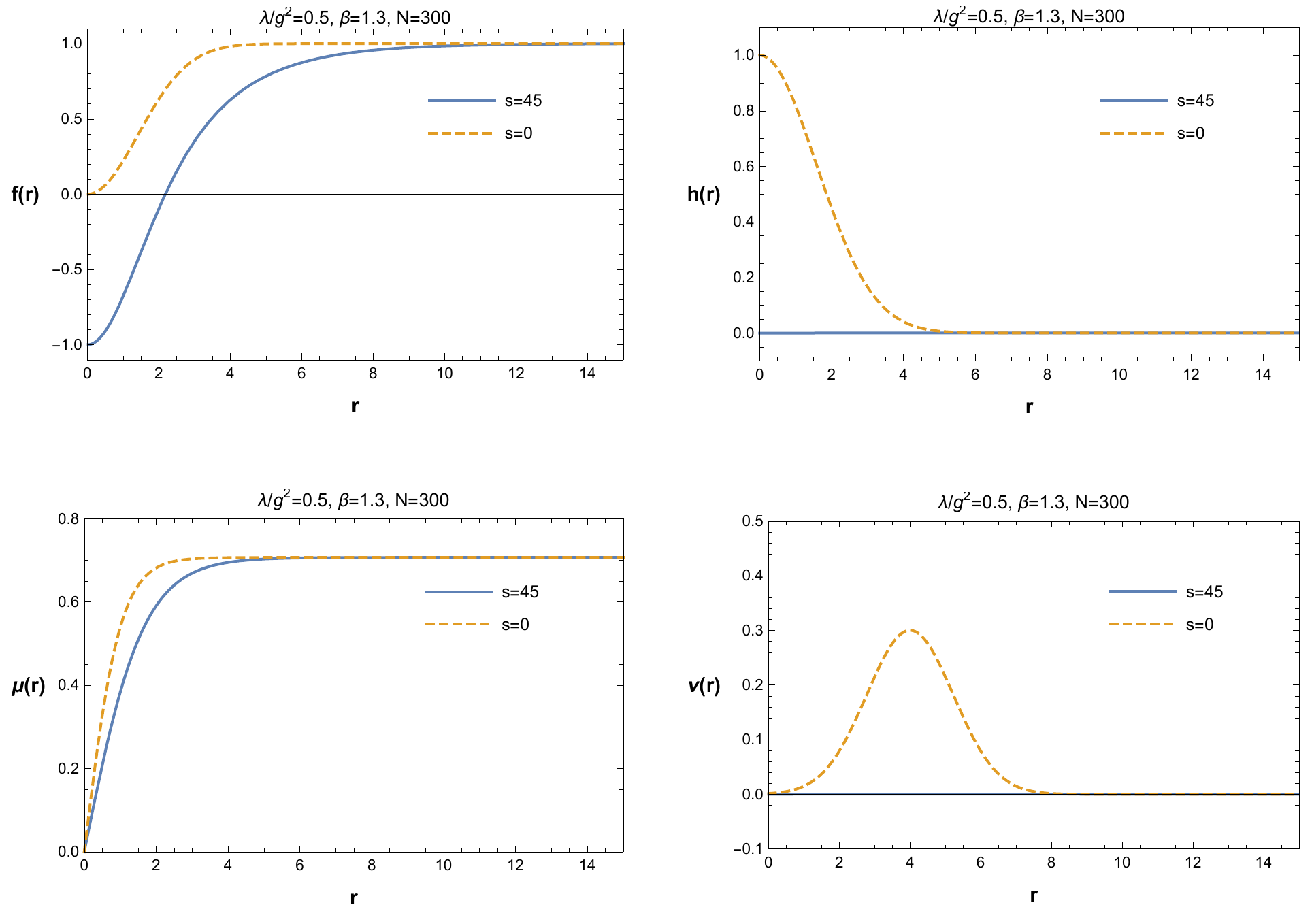}
 \caption{
The sphaleron configuration obtained by our flow equation.
 The solid blue lines are the flowed configurations at $s=45$,
 and the dashed orange lines are initial ones at $s=0$.
Top-left panel : $f(r)$, 
top-right panel : $h(r)$, 
bottom-left panel : $\mu(r)$,
bottom-right panel : $\nu(r)$.
 }
 \label{115217_8Dec19}
\end{figure}

\begin{figure}[htbp]
 \centering
 \includegraphics[width=0.6\textwidth]{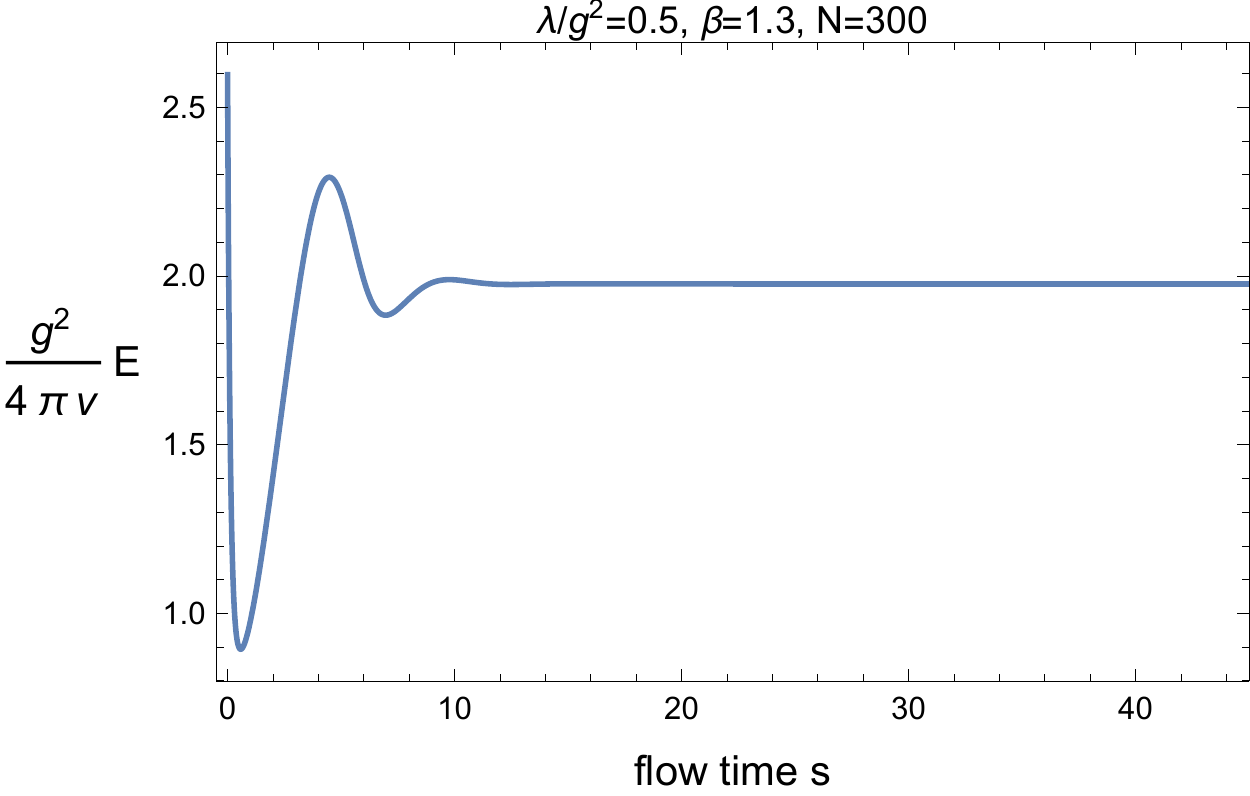}
 \caption{The evolution of energy $E$ with increasing flow time $s$ ($ 0\leq s \leq 45$).
 The plot shows the value of the energy divided by $4\pi v /g^2$ for convenience.
 The energy converges at $s\sim 10$.
 }
 \label{151140_8Dec19}
\end{figure}

\begin{figure}[htbp]
 \centering
 \includegraphics[width=0.6\textwidth]{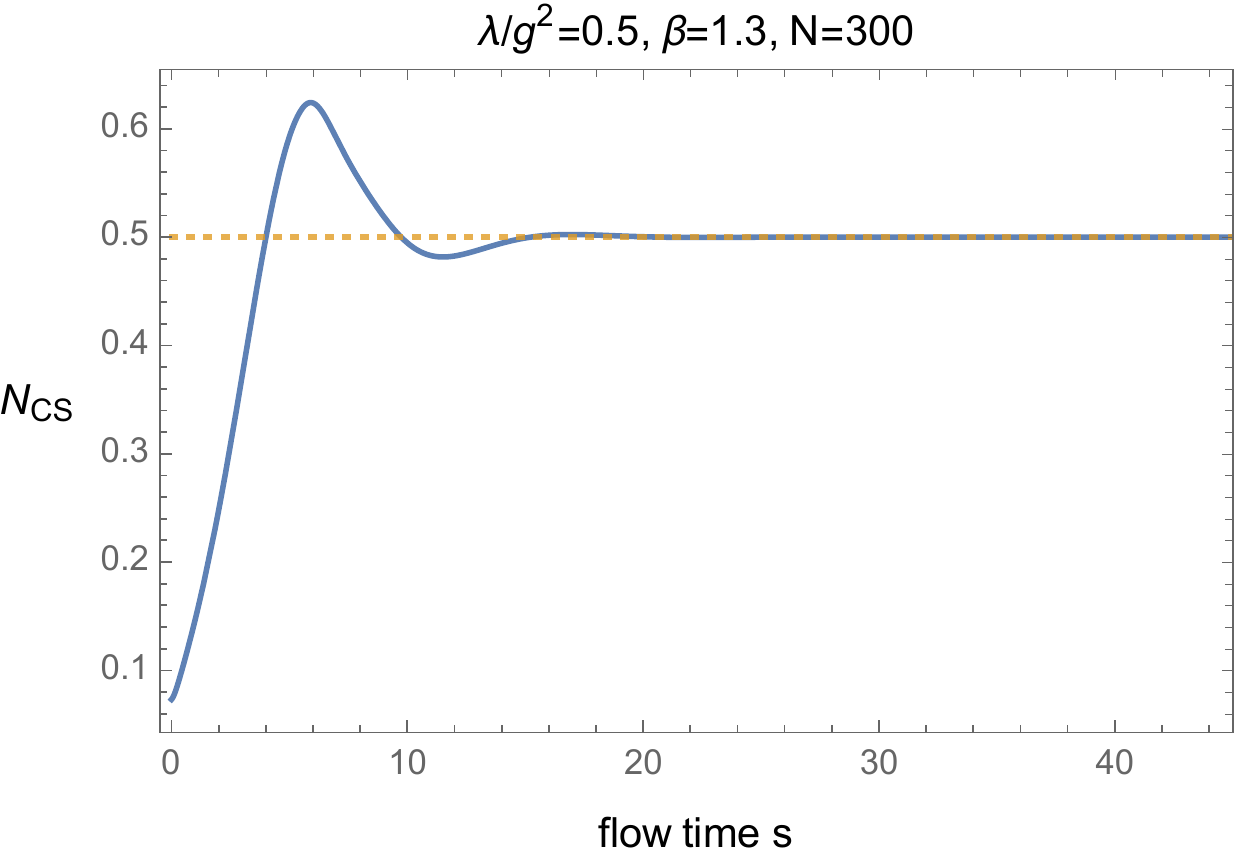}
 \caption{The solid blue line is the evolution of the Chern-Simons number with increasing flow time $s$ ($ 0\leq s \leq 45$).
 The dotted orange line is $N_{CS}=1/2$.
 The blue line clearly converges to $1/2$.
 }
 \label{151152_8Dec19}
\end{figure}

As a benchmark point, we take $\lambda/g^2=0.5$.
In addition, we also take $N=300$ and $\beta=1.3$.
We evolve nonspecial initial configurations at $s=0$ by the flow equations \eqref{maineq1} and \eqref{maineq2}.
The flowed result at $s=45$ is shown in Fig.~\ref{115217_8Dec19}.
The solid blue lines are flowed configurations at $s=45$.
The dashed orange lines are initial ones at $s=0$,
which are taken to satisfy the boundary conditions Eq.~(\ref{184233_9Dec19}).
The profiles of the flowed ones agree with the sphaleron
studied in the previous works \cite{Manton:1983nd,Yaffe:1989ms,Manton:2004tk}.
Furthermore, Fig.~\ref{151140_8Dec19} shows the evolution of energy for the flow time $s$
corresponding to the result in Fig.~\ref{115217_8Dec19}.
The energy oscillates for small $s$,
\footnote{
Note that the energy does not decrease monotonically in our flow equation
unlike the ordinary gradient flow
because of the modification term proportional to $\tilde{\beta}$.
}
but almost converges for $s\geq 10$.
The converged value is $E_{sph.}=1.976289 \times 4\pi v/g^2$,
which also agrees with that of in Refs.~\cite{Yaffe:1989ms,Manton:2004tk} in good precision.
In addition, Fig.~\ref{151152_8Dec19} shows the evolution of the Chern-Simons number,
\footnote{
The Chern-Simons number is calculated after moving to another gauge in which $\chi(0)=\chi(L)$ since it is not well defined in our gauge $a_0=a_1=0$.
}
which successfully converges to $1/2$.
These results show that we have obtained the sphaleron based on the gradient flow equations (\ref{maineq1}) and (\ref{maineq2}) .

We emphasize that the imaginary parts $h(r)$ and $\nu(r)$ are not fixed to $0$
as done in Refs.~\cite{Manton:1983nd,Yaffe:1989ms}.
Instead, they automatically decrease to $0$ along the flow.
Therefore, the flow is stable around the sphaleron
as we stated in Sec.~\ref{sec:formulation}.


We also investigate the sphalerons for several choices of $\lambda/g^2$.
We take $N=300$ and $\beta=1.3$ for $\lambda/g^2 \leq 5$, $\beta=5$ for $\lambda/g^2 = 10, 13$, respectively.
We evolve initial configurations using the flow equation until $s=45$.
Table~\ref{155438_8Dec19} shows the dependence of energy of the sphaleron on $\lambda/g^2$.
All the obtained energy values in Table~\ref{155438_8Dec19} are consistent with
the previous studies \cite{Yaffe:1989ms,Manton:2004tk}.
This shows that our flow equation works well for a parameter range of $\lambda/g^2$.

\begin{table}[htbp]
 \centering
\begin{tabular}{c|c}
 $\lambda/g^2$ & $E_{sph.} g^2/4\pi v$ \\
 \hline 
 0.001 & 1.56254 \\
 0.01 & 1.644148 \\
 0.5 & 1.976289 \\
 1 & 2.065615 \\
 5 & 2.28022 \\
 10 & 2.36486 \\
 13 & 2.39453
\end{tabular}
\caption{
The energy of the sphaleron obtained by our flow equation
for various choices of $\lambda/g^2$.}
\label{155438_8Dec19}
\end{table}

\subsection{Discussion}
As we have shown above, our results agree well with the previous studies.
This strongly suggests the validity of our method to obtain the sphaleron.
Especially, we emphasize that we did not choose any specific initial configurations.
In other words, the computations of the flow do not need any fine-tuning.
Instead, the field configuration automatically converges to the sphaleron along the flow.
This is an advantage of our method
compared to the previous ones in Refs.~\cite{Manton:1983nd,Yaffe:1989ms,Manton:2004tk}.

There are few cases that do not flow to the sphaleron but to the vacuum, 
although the latter is an unstable fixed point in our flow equations \eqref{maineq1} and \eqref{maineq2}.
For instance, consider $f(r)=\mu(r)/gv=\tanh(rv)$ and $h(r)=\nu(r)=0$ at $s=0$.
In this case, the flow equation
reduces to the ordinary gradient flow equation,
which results in a vacuum,
because the modification term vanishes for $h(r)=\nu(r)=0$ at $s=0$.
In other words, such initial configurations are ``fine-tuned'' to flow to the vacuum.
However, such a problem is easily avoided by starting from initial configurations
such as $h(r)\neq 0$.

In addition, note that the stability of the sphaleron is ensured 
only for configurations in the vicinity of the sphaleron.
Thus, when we start at a point too far from the sphaleron,
it cannot flow to any fixed points but diverge.

\section{Summary} 
\label{sec:summary}
We have given a simple formalism to obtain the sphaleron configuration
in gauge-Higgs theories using the gradient flow.
Based on the work in Ref.~\cite{Chigusa:2019wxb},
where the bounce configurations are discussed, we propose a new method to obtain the sphaleron solutions.
The most important point is how to choose the modification term appropriately.
In this paper, we have given the term as a gradient of the Chern-Simons number with respect to fields.
The modified flow equation in this manner gives the sphaleron configuration
as a stable fixed point of the flow in the large flow time.
Applying the methods to the $SU(2)$-Higgs model in $(3+1)$ dimensions,
which is well known for the existence of the sphaleron,
we give the modified flow equation of the $SU(2)$-Higgs model.
Solving the equation numerically,
we indeed obtain the sphaleron configuration that agrees with previous studies in a good precision.
We have also shown that the Chern-Simons number of the configuration converges to a half-integer.
Those results strongly suggest the validity of our method to obtain the sphaleron
without any fine-tuning of the initial configurations. 

Our method would also be applicable to the sphalerons
in the $SU(2)_W \times U(1)_Y$ Weinberg-Salam theory with $\theta_W\neq 0$.
In the model, there are many previous studies of the sphaleron
\cite{Manton:1983nd,Klinkhamer:1984di,Kunz:1992uh,Klinkhamer:1990fi,James:1992re}.
However, the detailed properties are still unclear 
except for the case of the small Weinberg angle, $\theta_W \simeq 0$.
Especially, a relation between the sphaleron and the Nambu monopole pair \cite{Nambu:1977ag}
has been argued in Refs.~\cite{Manton:1983nd,Barriola:1993fy,Vachaspati:1994ng,Hindmarsh:1993aw} 
(for a review, see \cite{Achucarro:1999it}), 
but remains unanswered.
In addition, it is also interesting to consider the sphalerons in extensions of the Higgs sector of the SM.
For instance, two Higgs doublet models (2HDM), in which an additional Higgs doublet is added into the SM (for reviews, see, e.g., Refs.~\cite{Gunion:1989we,Branco:2011iw}), 
have the sphaleron solution \cite{Grant:2001at,Kastening:1991nw,Moreno:1996zm}, 
although it is less known than that of the SM.
The relation with the Nambu monopole in 2HDM \cite{Eto:2019hhf,Eto:2020hjb} is also unclear.

It is also an interesting future work to consider the deformed sphaleron.
We have considered the sphaleron in the $SU(2)$-Higgs theory
with $10^{-3} \leq \lambda/g^2 \leq 13$ in this paper.
For $\lambda/g^2 > 18.1$, however,
a new sphaleron solution called the deformed sphaleron appears
\cite{Yaffe:1989ms,Kunz:1988sx},
whose Chern-Simons number is not half-integer.
As we stated above, such a solution cannot be obtained by our choice for $\mathcal{G}_A$, Eq.~\eqref{181134_27Nov19};
thus, we have to reconsider the flow equation in order to obtain the deformed sphalerons.

\section*{Acknowledgments}

We would like to thank So Chigusa and Yutaro Shoji for useful discussions and explanations of their work.
We also thank Masanori Tanaka for useful discussions.
This work is supported by JSPS KAKENHI Grants No.~JP18K13546 (K. K.) and No.~JP18J22733 (Y. H.).
We thank Iwanami Fujukai for the financial aid.

\bibliographystyle{utphys}
\bibliographystyle{unsrt}

\providecommand{\href}[2]{#2}\begingroup\raggedright\endgroup


\end{document}